\documentclass[11pt]{article}

\usepackage[a4paper,margin=1in]{geometry}

\usepackage[T1]{fontenc}

\usepackage[utf8]{inputenc}

\usepackage{amsmath,amssymb}

\usepackage{graphicx}

\usepackage{booktabs}

\usepackage{array}

\usepackage{caption}

\usepackage[numbers,sort&compress]{natbib}

\usepackage[hidelinks]{hyperref}

\usepackage{xcolor}

\usepackage{setspace}

\hypersetup{
  pdftitle={Network architecture shapes communication and computational function in simulated neurons-on-a-chip},
  pdfauthor={Michael Taynnan Barros},
  pdfsubject={Cell Systems Article},
  pdfkeywords={neuronal biocomputing, neurons-on-a-chip, neuromorphic engineering, network architecture, spiking neural networks, reservoir computing}
}

\newcommand{\etaSF}{\eta_{\mathrm{SF}}}

\definecolor{goldnote}{HTML}{9A6B00}

\definecolor{qcPurple}{HTML}{6F2DA8}

\definecolor{qcYellow}{HTML}{C49A00}

\newcommand{\rededit}[1]{{\color{red}#1}}
\definecolor{ntOrange}{HTML}{D97706}

\title{From communication to computation in neurons-on-a-chip: an \textit{in silico} study of neurotopomorphic computing}

\author{Michael Taynnan Barros$^{1,*}$\\[4pt]
$^{1}$School of Computer Science and Electronic Engineering,\\
University of Essex, Colchester, UK\\
$^{*}$Correspondence: m.barros@essex.ac.uk\\
ORCID:0000-0002-9765-7660}

\date{}

\begin{document}

\maketitle

\noindent\textbf{Article type:} Article

\vspace{6pt}


\section*{Highlights}

\begin{itemize}
    \item Neurons-on-a-chip networks produce distinct communication phenotypes.
    \item Networks scoring highest on the new IC$^3$ framework were among the poorest decoders.
    \item Simple feedforward chains outperformed densely connected neuronal networks.
    \item Results motivate neurotopomorphic computing through engineered network organisation.
\end{itemize}

\section*{Summary}
Living neuronal networks transform inputs through recurrent cellular and population dynamics, yet it is unknown which network architecture supports which computation. Neurons-on-a-chip turn this question into a design problem because microchannels guide axonal growth and set the network architecture. We introduce IC$^3$, an Integrated Characterisation of Communication-Driven Computation, which characterizes network state through neuronal dynamics, functional communication, and structural support. We implemented nine architectures \textit{in silico} as conductance-based spiking networks and tested each on frequency decoding, temporal-order discrimination, and fading memory. Predominantly feedforward circuits decoded best. Sequential Chain and Microchannel Diode had the lowest IC$^3$ and recruited a third of reachable neurons, yet achieved the two highest scores on both classification tasks. Across architectures, higher IC$^3$ went with lower classification scores. We term this new direction \emph{neurotopomorphic computing}, in which the physical organisation of neuronal connectivity is engineered as part of the computing substrate.

\section{Introduction}

Living neuronal networks can transform structured stimulation into population responses that support task-relevant readouts \cite{Habibollahi2023,Sumi2023}. Closed-loop interaction can further modify neuronal activity during task performance, and structured input can reorganise their collective dynamics \cite{Kagan2022,Habibollahi2023}. With a trained readout attached, cultured networks have separated input patterns and, under feedback control, generated target temporal signals \cite{Sumi2023,Sono2026}. Two questions remain open\rededit{~\cite{Habibey2022,Buentello2024,Loeffler2026}}: how does neuronal architecture organise the communication state expressed by a neuronal network, and how does that state relate to the computational transformations accessible to a readout?

In neurons-on-a-chip, the physical layout of the circuit is a design variable for its network. Cell placement, axonal growth, stimulation, and electrophysiological recording can all be controlled within the same culture system \cite{Buentello2024,Maoz2021}. Because these platforms can incorporate human-derived cells, they are also used to study disease mechanisms, neurotoxicity and therapeutic response \cite{Buentello2024,Maoz2021}. The question is whether different cultures with the same neuronal layout produce similar responses to the same stimulation \cite{Sato2023,Habibey2024}. Activity has to follow the intended routes and still carry usable information when it reaches downstream neurons. Microfabrication sets the distances, bottlenecks, directional pathways and recurrent loops along which this may be engineer to happen \cite{Habibey2022,Gladkov2017,Habibey2024}.

Several fabrication strategies control neuron-on-a-chip properties at different spatial scales, and each alters the resulting network dynamics \cite{Montala2022,Habibey2024}. Adhesive patterning can confine small neuronal populations to isolated microislands, and topographical tracks and valleys constrain neurite growth and organise collective activity \cite{Walker2021,Montala2022}. Compartmentalised microfluidic devices separate neuronal somata into defined modules and connect them through axon-guiding microchannels, so that modular circuits can be assembled and recorded on microelectrode arrays \cite{Habibey2022,SheinIdelson2011,Sato2023,Habibey2024}. Channel geometry can also impose directionality, e.g. asymmetric microchannels and microtunnels bias axonal growth, and circular and pseudo-feedforward couplings create directional communication between modules \cite{Gladkov2017,Forro2018,Monma2025}. More recently, microscale scaffolds and single-cell printing have enabled reproducible human neuronal circuits with defined cell placement \cite{Striebel2025}.

We represent the engineered neuronal circuits using nine candidate network topologies, or architectures. Network architecture can be controlled experimentally in neurons-on-a-chip \cite{Habibey2022,Habibey2024}. However, comparing different architectures \emph{in vitro} requires fabricating each design and maturing separate cultures. Biological variation between cultures can then be difficult to separate from the effects of topology as well as be resource-intensive. We therefore compared the architectures \emph{in silico}, where topology can be changed with the neuronal model and stimulation fixed. Synthetic biology uses the same simulation-informed methodology when candidate circuits are compared computationally before construction \cite{Darlington2020}. Each of the nine architectures has an experimental or theoretical precedent. All-to-All is the fully connected reference \cite{Lopes2019}. Sparse and Directed Random are directed Erd\H{o}s--R\'enyi ensembles with connection probabilities of 0.10 and 0.12 \cite{Roxin2011}. Star routes all traffic through one hub, as in hub-centred micropatterns \cite{Burbulla2016}. Sequential Chain links populations one after another, as directed microchannels do \emph{in vitro} \cite{Habibey2022}. Microchannel Diode adds weak reverse access to the chain, after axonal-diode designs that bias neurite growth in one direction \cite{Habibey2022}. Clustered Ring couples dense modules in a cycle, following engineered modular circuits with controlled intermodule coupling \cite{Yamamoto2023,Monma2025}. Small-World and Scale-Free reproduce two organisations reported in cultures: local clustering with a few long-range shortcuts \cite{Downes2012}, and a heavy-tailed degree distribution with a few hubs \cite{Honegger2016}. Choosing between them needs measurements that link the imposed design to the neuronal state it produces and to what that state can compute. Structural connectivity specifies where signals could travel but not that neurons after many hops will be activated by the initial input signal \cite{Habibey2024,Honegger2016}. 

Transfer entropy (TE) usually estimates directed predictive information transfer between pairs of spike trains. However, TE does not represent the population states or structural constraints within which that transfer occurs \cite{Schreiber2000}. Task performance is a joint test of the input, the activity it evokes, the features extracted from that activity and the readout. Architecture, activity and task output are therefore three levels of one engineered system, and each needs its own characterization \cite{Schreiber2000,Dambre2012,Habibey2024}. A useful design framework must connect these levels without treating network characterization as equivalent to computational performance \cite{Habibey2022,Habibey2024,Dambre2012,Loeffler2026}.

Our previous work treats neuronal circuits as communication systems that can be engineered to compute. We first studied engineered
neuron-astrocyte signalling and synthetic neuronal logic circuits as
information-processing systems \cite{Adonias2020,Barros2021}. We then extended
this direction towards reservoir computing and sequential neuronal
biocomputation, examining how neuronal dynamics and circuit organization shape
computational behaviour \cite{Schoefmann2024,Basso2026}. Our more recent work found that communication and computation differ with topology in bioengineered networks, and proposed neurons-on-a-chip as a platform for engineering biological intelligence by design \cite{Barros2024,Mullett2026}. Loeffler et al. have since shown in living cultures that microengineered architecture changes both electrophysiological dynamics and decoding performance \cite{Loeffler2026}.

Here we introduce IC$^3$, an Integrated Characterisation of Communication-Driven Computation, to link the state a network enters under stimulation to what it can compute. IC$^3$ splits that state into three coupled domains. \emph{Neuronal dynamics} covers the activity states the network generates, \emph{functional communication} the information exchanged as activity propagates, and \emph{structural support} the architecture that constrains both. For neurons-on-a-chip we integrate IC$^3$ with eight network observables: spike entropy, effective dimensionality, inter-spike-interval variability, burst-like and regular-spiking fractions, mutual information, TE and a graph-structure.

We apply IC$^3$ in four steps. We first measure how architecture and stimulation frequency change IC$^3$ and TE, and how stable the IC$^3$ ranking is to its weights. We then compare graph reachability with the neurons that actually respond, and path delay with response latency. From response probability, firing rate, latency, timing variability, information transfer and spike expenditure we build each network's \emph{communication phenotype}. Finally, we test every network on close-frequency decoding, temporal-order discrimination, and fading memory.  

\begin{figure}[p]
    \centering
    \includegraphics[width=\textwidth]{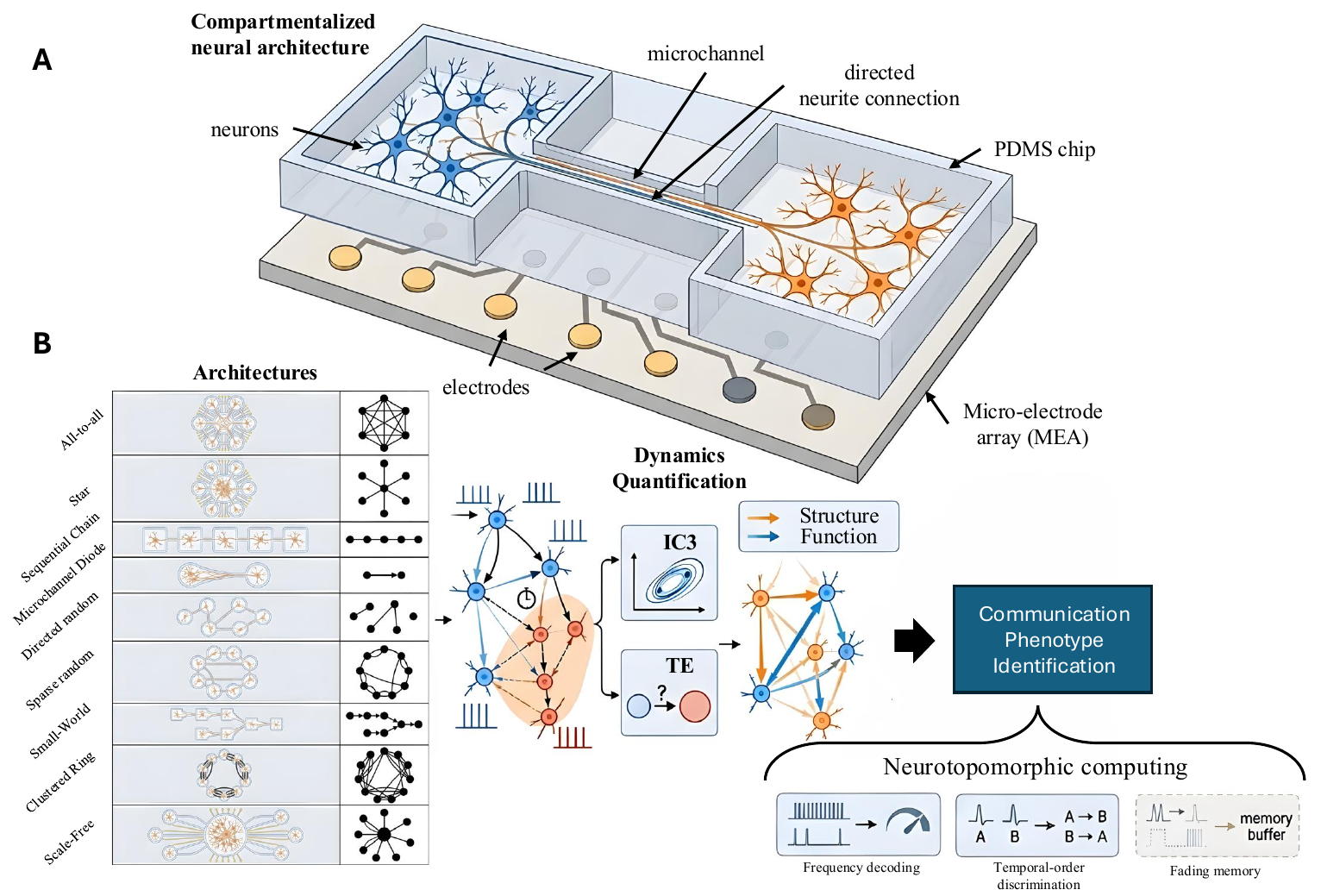}
    \caption{\textbf{Neurons-on-a-chip architecture links physical circuit design to neuronal dynamics, functional communication, and computational function.}
\textbf{a}, Conceptual neurons-on-a-chip platform integrating patterned neuronal networks, PDMS microchannels and compartments, extracellular recording and stimulation electrodes, and an MEA substrate.
\textbf{b} Nine candidate architectures were evaluated as graph-defined neuronal circuits. Their activities were characterized using IC$^3$, which integrates neuronal dynamics, functional communication, and structural support, together with transfer entropy and propagation-related measurements. The resulting network states were then assessed using close-frequency decoding, temporal-order discrimination, and fading-memory tasks.}
    \label{fig:overview}
\end{figure}

Across nine architectures and 20 independently seeded network instances per architecture, topology produced distinct communication states and task-dependent computational profiles. Sequential Chain and Microchannel Diode had the lowest IC$^3$ yet produced the two highest mean scores for frequency decoding and temporal-order discrimination, and fading-memory performance remained near baseline in every architecture. Across architecture means, higher IC$^3$ and higher TE were associated with lower classification scores ($\rho$ between $-0.73$ and $-0.81$). These results motivate what we term \emph{neurotopomorphic computing}, in which the physical wiring of a neuronal circuit is designed alongside its stimulation, dynamics and readout. A candidate circuit can then be screened in simulation, for the communication state it produces and for whether that state supports the intended computation, before it is fabricated.

\section{Results}

\subsection{IC$^3$ quantifies architecture-dependent communication--computation configurations}

We quantified how architecture and input jointly changed the IC$^3$ configuration. The analysis contained nine architectures with 20 independently seeded network instances. Each 15-neuron unit received inputs at 0.5, 1, 2, 5, 10, 20, and 40\,Hz, with three stochastic trials per frequency.

Architecture, stimulation frequency, and their interaction affected IC$^3$. Mean IC$^3$ was highest for All-to-All (0.354, 95\% CI 0.344--0.364), Small-World (0.344, 0.324--0.363), and Scale-Free (0.338, 0.318--0.357) networks. Sequential Chain (0.154, 0.120--0.187) and Microchannel Diode (0.164, 0.128--0.201) networks had the lowest values.

\begin{figure}[p]

    \centering

    \includegraphics[width=0.6\textwidth]{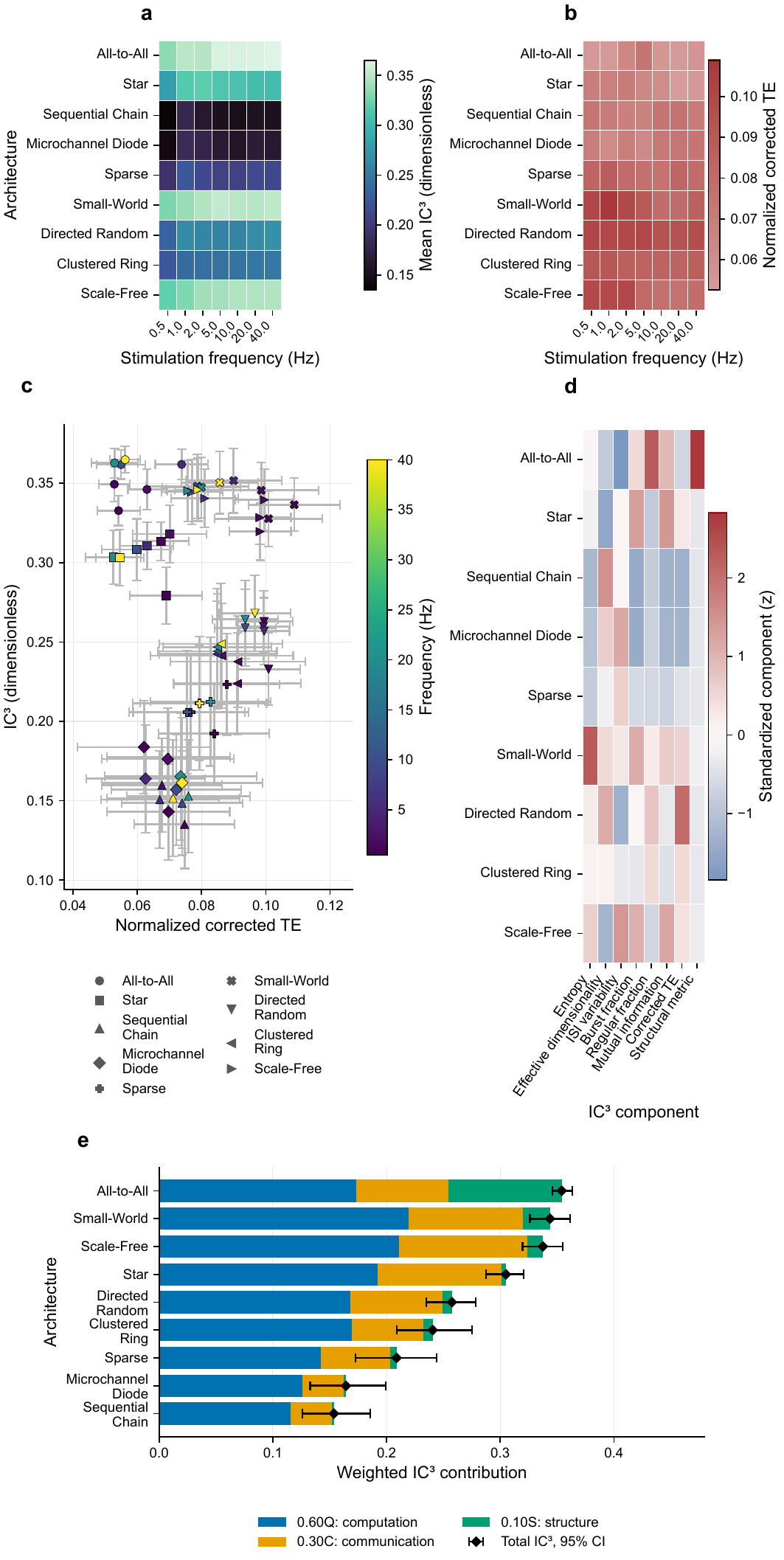}

    \caption{\textbf{IC$^3$ captures architecture-dependent network configurations.}
\textbf{a}, Mean IC$^3$ values for the nine neuronal architectures across seven stimulation frequencies.
\textbf{b}, Mean target-entropy-normalized, shuffle-corrected transfer entropy (TE) across the same architecture and frequency sweep.
\textbf{c}, Joint IC$^3$--TE state space for architecture and frequency means. Markers identify architectures, colours indicate stimulation frequency, and error bars show 95\% confidence intervals across network instances.
\textbf{d}, Standardized values of the eight IC$^3$ components across architectures, showing how neuronal dynamics, functional communication, and structural support contribute differently to each network configuration.
\textbf{e}, Decomposition of mean IC$^3$ into neuronal dynamics, functional communication, and structural support. Stacked bars show the weighted domain contributions with mean IC$^3$ with 95\% confidence intervals across network instances.}

    \label{fig:claim1}

\end{figure}

TE was also architecture- and frequency-dependent, but it produced a different ordering. Directed Random networks had the highest target-entropy-normalized  TE (0.0975, 95\% CI 0.0887--0.1064), followed by Small-World networks (0.0918, 0.0771--0.1065). Under the present IC$^3$ operationalization, All-to-All networks combined the highest scalar IC$^3$ with the lowest mean TE (0.0583, 0.0534--0.0631). TE is one of the eight IC$^3$ components (weight 0.15), so the two measures are not independent. Their different orderings arise from the interaction of TE with the other seven components (Figure~\ref{fig:claim1}c).

IC$^3$ architecture rankings were unchanged across 500 local weight perturbations, so the rankings were insensitive to local changes in the weights. The domain decomposition showed that similar scalar IC$^3$ values arose from different configuration profiles (Figure~\ref{fig:claim1}e). Under the present scaling, All-to-All received a structural-support contribution of 0.100 and a combined neuronal dynamics and communication contribution of 0.254. Small-World and Scale-Free received structural contributions of 0.024 and 0.014 and combined neuronal dynamics and communication contributions of 0.320 and 0.324, respectively.

\subsection{Different architectures produce distinct communication phenotypes under tested stimulation conditions}

\begin{figure}[p]

    \centering

    \includegraphics[width=0.9\textwidth]{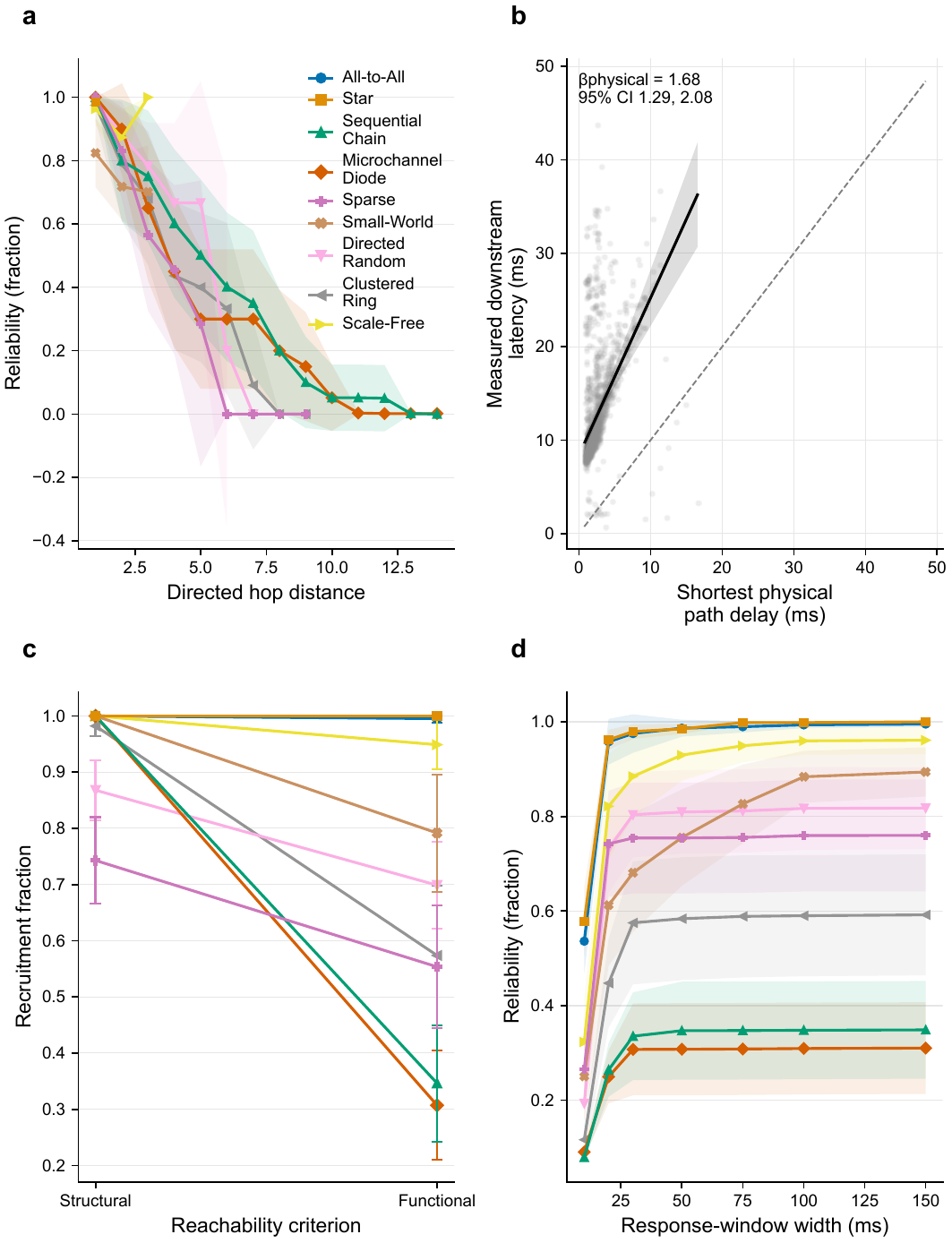}

    \caption{\textbf{Architecture shapes the conversion of structural connectivity into downstream neuronal responses.}
\textbf{a}, Mean post-stimulus response probability as a function of network hop distance, with 95\% confidence intervals across independently seeded network instances.
\textbf{b}, Relationship between post-stimulus first-spike latency and the shortest physical path delay from the stimulated source to the responding neuron. 
\textbf{c}, Architecture-level comparison between the fraction of downstream neurons that were structurally reachable from the stimulated source and the fraction that responded after at least 50\% of stimulus events ($R\geq0.5$).
\textbf{d}, Mean post-stimulus response probability across alternative response-window durations. Architecture-specific differences remained across the tested windows, while response probability approached a plateau at longer windows.}

    \label{fig:claim2}

\end{figure}

We next examined how architecture shaped the functional
expression of these pathways through downstream post-stimulus recruitment and
propagation. To quantify this, we applied a 5\,Hz, 50\%-duty-cycle square-wave current with an amplitude of 6 model units to one source neuron for 2000\,ms. Three stochastic trials were performed for each architecture--network instance. A neuron was structurally reachable when a directed path connected it to the stimulated source. Post-stimulus response probability was defined as the fraction of stimulus onsets followed by at least one target spike within the response window, and functional recruitment was defined at $R\geq0.5$. We measured both quantities for every downstream neuron, excluding the directly stimulated source (Figure~\ref{fig:claim2}).

Post-stimulus response probability declined with hop distance, and the rate of decline depended on architecture. The decline was not monotonic in every architecture, since Scale-Free and Directed Random networks rose at intermediate hops (Figure~\ref{fig:claim2}a). Cluster-robust analysis identified effects of hop distance, architecture, and their interaction (Figure~\ref{fig:claim2}a). The response-probability change per hop ranged from $-0.062$ for Small-World networks (95\% CI, $-0.114$ to $-0.010$) to $-0.175$ for Sparse networks ($-0.207$ to $-0.143$).

Shortest physical-path delay was also associated with post-onset first-spike latency. After adjustment for architecture, each additional millisecond of shortest-path transmission delay was associated with a 1.684\,ms increase in measured response latency (95\% CI, 1.287--2.082; $p<0.001$; Figure~\ref{fig:claim2}b). Because the slope exceeds one, latency grew faster than path delay, by about 0.68\,ms for each millisecond of delay. We hypothesise that this excess reflects integration at each successive relay. Depending on architecture, measured responses occurred 8.03--12.96\,ms later than predicted from transmission time alone. This residual interval is compatible with contributions from synaptic transmission, membrane integration, recurrent dynamics, stochastic timing, and numerical discretization.

Structural reachability overstated functional recruitment. With the predefined criterion $R\geq0.5$, the reachable fraction exceeded the recruited fraction by 0.264. They were closely matched in All-to-All (1.00 versus 0.995), Star (1.00 versus 1.00), and Scale-Free networks (1.00 versus 0.949). In Sequential Chain and Microchannel Diode networks, every downstream neuron was structurally reachable, whereas the functionally recruited fractions were 0.346 and 0.307, respectively. Clustered Ring networks showed an intermediate reduction from 0.982 structural reachability to 0.574 functional recruitment.

We then varied the response window from 10 to 150\,ms to test whether the low response probabilities resulted from counting only rapid spikes. Post-stimulus response probability initially increased with window duration and then approached a plateau, while architecture-specific differences remained (Figure~\ref{fig:claim2}d). At 150\,ms, mean response probability reached 1.00 for Star and 0.995 for All-to-All networks, but remained at 0.349 for Sequential Chain and 0.310 for Microchannel Diode networks.

\subsection{Neuronal architectures differ in recruitment, timing and reliability}

\begin{figure}[p]

    \centering

    \includegraphics[width=0.9\textwidth]{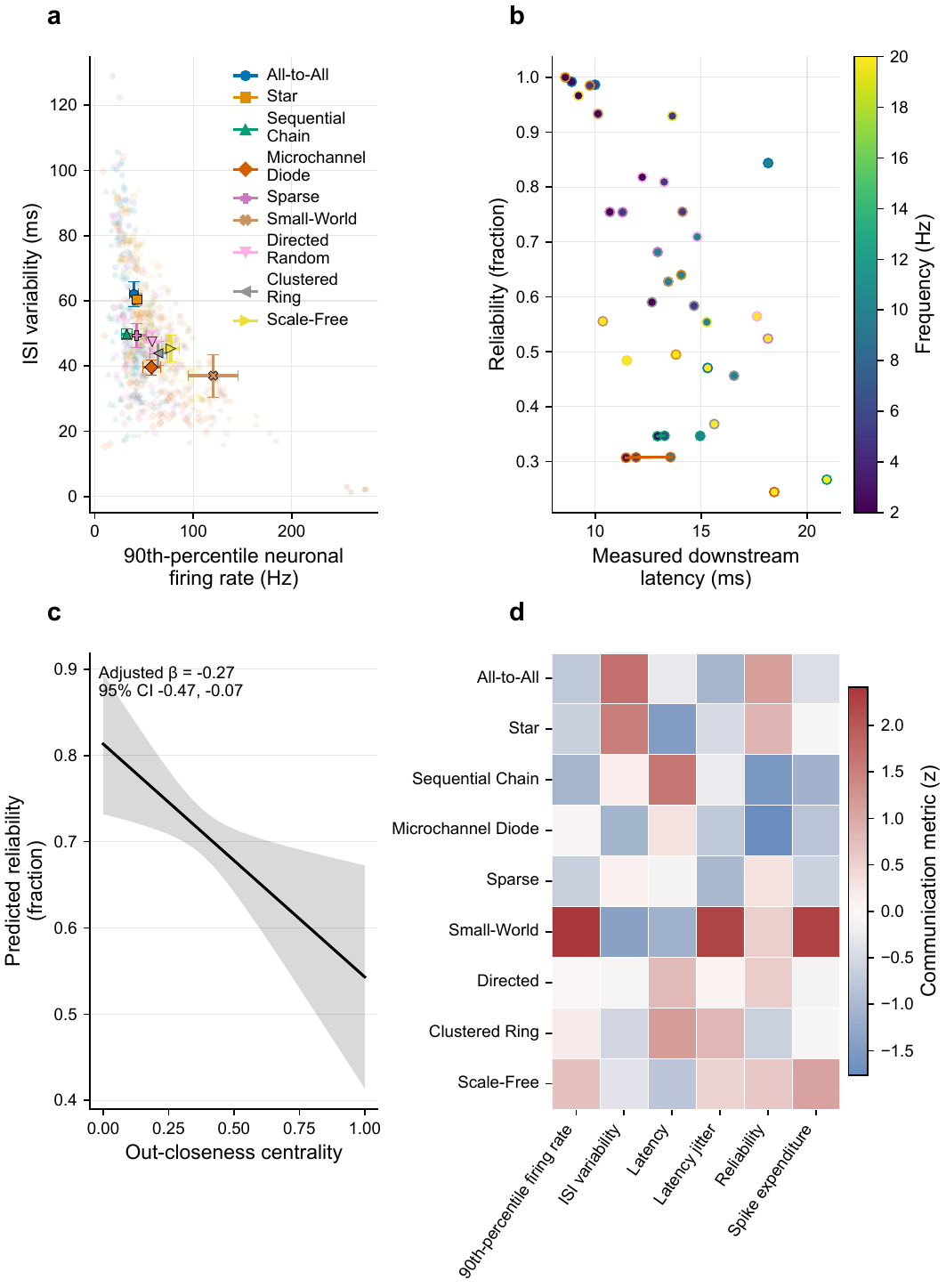}

    \caption{\textbf{Architecture-dependent communication phenotypes and their functional trade-offs.}
\textbf{a}, Relationship between upper-tail downstream firing rate and ISI variability across architectures and stimulation frequencies. 
\textbf{b}, Operating points relating post-stimulus response probability and response latency across stimulation frequencies. 
\textbf{c}, Model-predicted post-stimulus response probability as a function of out-closeness, evaluated at a hop distance of two and averaged equally across architectures. 
\textbf{d}, Standardized communication-phenotype profiles for each architecture across the measured communication endpoints. }

    \label{fig:claim3}

\end{figure}

For each independent network, we characterized its communication phenotype by measuring how strongly and reliably activity propagated through the circuit. We quantified this using the 90th-percentile neuronal firing rate, post-stimulus response probability, downstream latency, latency jitter, inter-spike-interval (ISI) variability, and spike expenditure across the matched input sweep (Figure~\rededit{\ref{fig:claim3}}). All six communication measures varied with architecture and stimulation frequency.

The multivariate profiles revealed architecture-specific patterns of functional communication (Figure~\ref{fig:claim3}a,b,d). Small-World networks had the highest mean 90th-percentile rate (120.3\,Hz) and the greatest standardized spike expenditure. Star networks combined high post-stimulus response probability (0.780) with the shortest mean downstream latency (11.55\,ms). All-to-All networks reached the highest response probability (0.823) at 13.09\,ms. Sequential Chain and Microchannel Diode networks used less activity but had lower response probabilities (0.327 and 0.292). Sequential Chain networks also had the longest mean latency (15.52\,ms). No architecture simultaneously maximized firing activity, recruitment, temporal precision, and activity efficiency.

We quantified each downstream neuron's network centrality using out-closeness, which measures how directly that neuron can reach other nodes through outgoing paths. After adjustment for hop distance and architecture, greater centrality was associated with lower post-stimulus response probability (95\% CI $-0.473$ to $-0.067$; Figure~\ref{fig:claim3}c). Centrality was not associated with firing rate (95\% CI $-9.37$ to 26.82).

Figure~\ref{fig:claim3}b,d shows the resulting design constraint. Across the nine architectures, a gain in one communication measure came with a loss in another, so architecture has to be chosen together with input frequency and the phenotype a device needs. A topology that maximizes propagation may be unsuitable when latency or spike expenditure is the limiting device constraint.

\subsection{Architecture shapes computational utility}

\begin{figure}[p]

    \centering

    \includegraphics[width=0.75\textwidth]{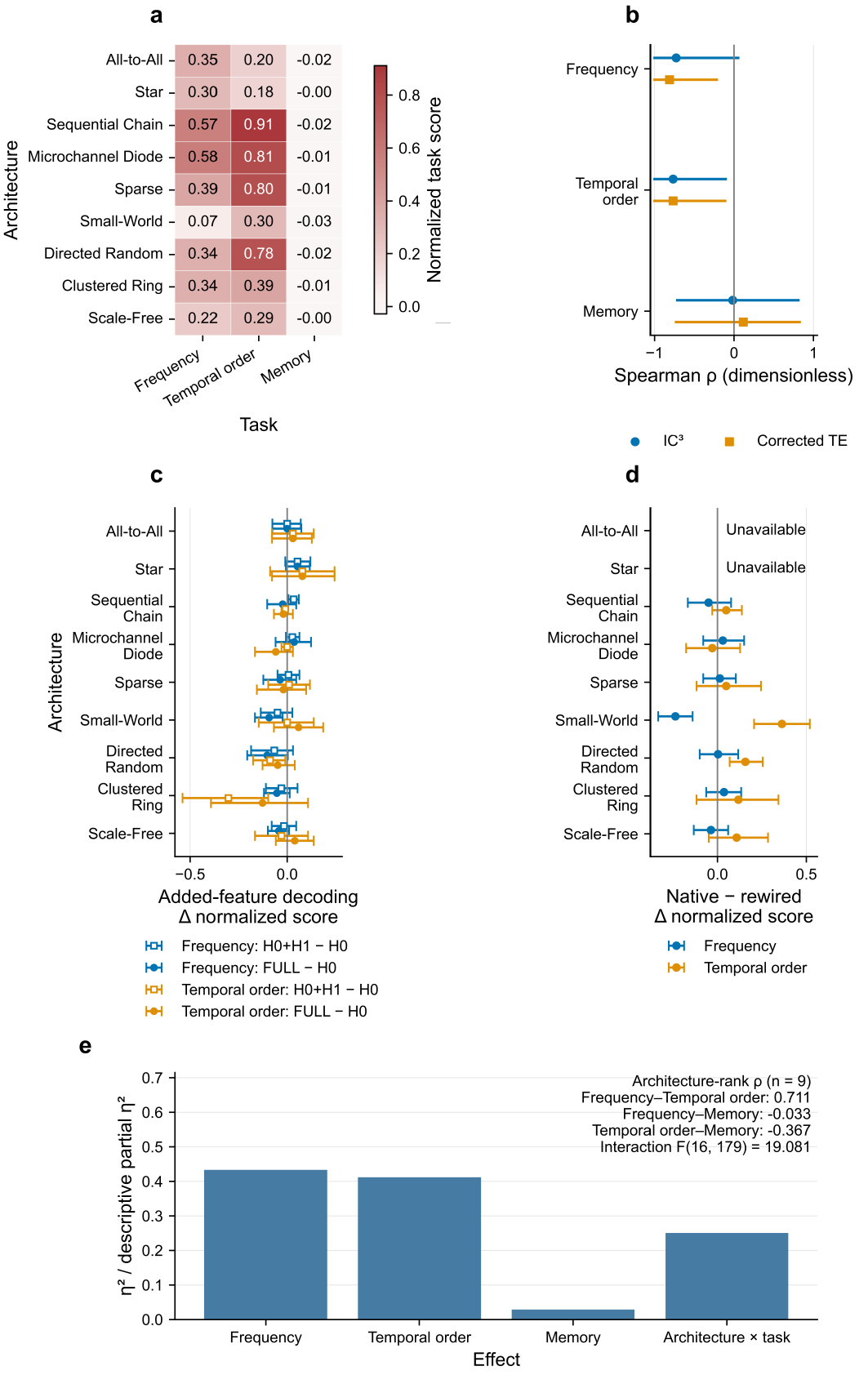}

    \caption{\textbf{Architecture-dependent computational performance and its relationship to network characterization.}
\textbf{a}, Mean normalized performance of the nine neuronal architectures across the three computational tasks: close-frequency decoding, temporal-order discrimination, and fading memory.
\textbf{b}, Spearman correlations across architecture means ($n=9$) relating task performance to IC$^3$ and transfer entropy (TE).
\textbf{c}, Change in frequency- and temporal-order decoding when the readout is expanded from the stimulated neuron(s) to include first-hop neurons or the full network.
\textbf{d}, Change in normalized task performance between native and degree-preserving rewired networks for frequency and temporal-order decoding; positive values indicate higher performance in the native architecture.
\textbf{e}, Strength of the architecture effect for each task ($\eta^2$), the descriptive partial $\eta^2$ for the architecture-by-task interaction, and Spearman correlations between architecture rankings across tasks ($n=9$).}

    \label{fig:claim4}

\end{figure}

We next tested whether each IC$^3$ configuration produced activity that supported a defined computational utility under the chosen feature representation and readout. Close-frequency decoding tested whether the network separated inputs at 8.5, 10.0, 11.5, and 13.0\,Hz. Temporal-order discrimination tested whether it distinguished A$\rightarrow$B from B$\rightarrow$A stimulation. The fading-memory task tested whether the network's current activity still contained information about inputs presented 25--200\,ms earlier. Classification trajectories were analysed with functionally aligned neuronal positions and nested regularization selection. Classification accuracy was normalized relative to task-level empirical chance. Fading-memory performance was assessed using held-out $R^2$. Added feature readouts and degree-preserving rewiring were analysed as sensitivity tests (Figure~\ref{fig:claim4}).

Architecture explained substantial
variation in both close-frequency decoding ($\eta^2=0.433$) and temporal-order
discrimination ($\eta^2=0.412$), while fading-memory performance showed little
architecture dependence ($\eta^2=0.029$). The architecture and task relationship
was also substantial (partial $\eta^2=0.251$), showing that the computational
effect of topology depended strongly on the transformation being performed.

Sequential Chain and Microchannel Diode produced the two highest mean scores for both frequency decoding and temporal-order discrimination. Across the nine architecture means, frequency and temporal-order rankings
were positively related ($\rho=0.711$), whereas frequency versus fading
memory ($\rho=-0.033$) showed little rank correspondence and temporal order
versus fading memory ($\rho=-0.367$) showed an inverse relationship. Fading-memory performance remained near baseline in every architecture. Cultured neuronal networks can exhibit
transient fading memory on subsecond timescales \cite{Dranias2013}, and repeated
stimulation can induce plastic changes that establish stimulus-specific memory
traces \cite{LeFeber2015}. The present result characterises untrained
fading-memory processing.

The relationship between network characterization and computation differed
between architecture-level and within-architecture analyses.
At the architecture level, IC$^3$ showed negative rank associations with both
supported classification tasks. This relationship was clearest for
temporal-order discrimination ($\rho=-0.767$), while frequency decoding showed
a similar negative ordering ($\rho=-0.728$).
TE showed comparable negative associations with frequency decoding
($\rho=-0.812$) and temporal-order discrimination ($\rho=-0.767$). Fading
memory showed little association with either IC$^3$ or TE.

Within architectures, the relationship between IC$^3$ and computation remained
task-specific. Temporal-order discrimination decreased as IC$^3$ increased
($\beta_W=-1.567$), supporting the same negative relationship observed across
architecture means. Frequency decoding showed a much weaker within-architecture
association ($\beta_W=-0.260$).

Expanding the readout beyond the directly stimulated neurons did not improve decoding in any tested architecture--task combination (Figure~\ref{fig:claim4}c). Rewiring effects were also architecture and task-specific.
Small-World networks favoured the rewired configuration for frequency decoding
($\Delta P=-0.237$), whereas Directed Random
($\Delta P=0.157$) and Small-World ($\Delta P=0.363$) favoured their native
configurations for temporal-order discrimination. Because rewiring also changes the network instance and, for some architectures, geometry and propagation delays, the contrast is interpreted at the level of the complete network configuration.

\section{Discussion}

Widespread network engagement did not coincide with the strongest computational performance in this study. All-to-All, Small-World, and Scale-Free networks occupied the upper end of the IC$^3$ range, whereas Sequential Chain and Microchannel Diode produced much lower IC$^3$ values. Yet the latter two architectures achieved the highest mean scores on both supported classification tasks. They also recruited only a fraction of the neurons that were structurally reachable from the stimulated source. This suggests that, in engineered neuronal systems, spreading activity across more of the network does not necessarily improve computation. Preserving input distinctions as signals move through the circuit may be more important.

The communication phenotype provides a functional description of how network architecture changes the response expressed under stimulation. Architecture shapes which neurons respond, when they respond, how consistent their timing is, how strongly they fire, and how much activity the response produces. These properties describe the neuronal state available to a readout without assuming that the communication phenotype itself mediates task performance. For neurotopomorphic computing, the aim is to relate an engineered network form to the communication state it produces and then ask which computational functions remain accessible under a defined readout.

The same view separates IC$^3$, TE and task performance. TE captures directed predictive information transfer between neuron pairs. By contrast, IC$^3$ describes a broader network state that also includes neuronal dynamics and structural support. Neither quantity increased with computational performance, and both were negatively rank-associated with the two classification tasks. Temporal-order discrimination decreased with increasing IC$^3$ at both the architecture and within-architecture levels, while frequency decoding showed a weaker within-architecture relationship. A network can therefore express a high-IC$^3$ state without making a particular task easier to solve. This distinction is consistent with reservoir-computing studies in which computational performance emerges from the interaction between input encoding, recurrent dynamics, and readout \cite{Dambre2012,Houben2026,Milisav2026}, and with work on spiking neuromorphic networks, in which capacity peaked at criticality but only complex tasks benefited \cite{Cramer2020}. IC$^3$ therefore characterises the state a network generates, and we do not use it to predict task performance. The name refers to the three domains IC$^3$ integrates. Task performance tests whether that state contains the distinctions a computation requires.

The readout analysis shows that extending the feature set from the directly stimulated neurons to first-hop neurons or the full network produced no supported improvement in decoding, despite the additional neurons contributing measurable activity. A larger feature set may contain more activity without containing more task-relevant variation, particularly when downstream responses are correlated with one another or with signals already present near the stimulation sites. It may also introduce variability that increases the apparent complexity of the representation without improving the stability of the decision boundary used by the readout. For the feature representation and readout used in our study, additional network activity did not provide additional information that improved held-out prediction. One possibility is that downstream responses become increasingly redundant with information already present close to the stimulation sites. Another is that propagation transforms task-relevant distinctions into forms that are less accessible to the chosen decoder. Our simulations suggest that information can remain present in a network while becoming less useful for a particular computation. We did not test the redundancy explanation directly.

We hypothesise that restricting how signals spread preserves the input distinctions these two tasks require. Sequential Chain and Microchannel Diode both constrain propagation predominantly in one direction. In living cultures, unidirectional axon routing between modules improved reservoir performance over bidirectional channels \cite{Habibey2022}. Across the nine architectures, more connectivity, greater recruitment, and stronger overall network activity did not consistently produce better task performance. However, our design does not isolate directionality. The two architectures also differ from the others in four respects that may contribute: i) path redundancy, ii) geometry, iii) propagation delay and iv) recurrent coupling. The circular embedding of the Microchannel Diode also placed the second temporal-order stimulation site at a different chain position from the Sequential Chain, so Chain--Diode differences cannot be attributed to directionality alone.

In an in vitro circuit, the connectivity that neurotopomorphic computing treats as a design variable is realised through five physical properties, namely directionality, compartmentalisation, path redundancy, path length and spatial routing, and each can alter how activity propagates. These properties can be patterned experimentally with cell source, stimulation and readout protocols. Communication and computation can then be compared across physically distinct circuits with everything else matched.

No architecture produced fading memory. Performance stayed close to baseline in all nine over lags of 25--200\,ms ($\eta^2=0.029$), although the same architectures separated close-frequency and temporal-order inputs strongly. Topology changed how current input propagated but had no measurable effect on memory of earlier input. We define input fidelity as the preservation of current-input distinctions at the readout, and persistence as the recovery of past input from the current state (held-out $R^2$ at lags of 25--200\,ms). We hypothesise that the two place different demands on recurrence, and none of the tested architectures achieved both. The null result is specific to the present network, synaptic, stimulation and readout configuration. Cultured neuronal networks do hold short-term memory and stimulus-specific traces under other protocols \cite{Dranias2013,LeFeber2015}.

Structural reachability set where activity could travel but did not determine which neurons were recruited. Response probability declined with hop distance on average, and downstream latency grew faster than physical path delay. After adjustment for hop distance and architecture, neurons with higher out-closeness responded to fewer stimulus onsets, so a structurally favourable position did not make a neuron a stronger functional relay.

We argue that the design target for neurons-on-a-chip is the transformation between stimulation and readout. Temporal-order discrimination, close-frequency decoding and memory place different demands on the measured neuronal response \cite{Loeffler2026,Habibey2024}. On a device, these demands map onto coupled variables that include microfluidic directionality, compartmentalisation, recurrent coupling, path length, stimulation protocol and readout location. Several can be fabricated before culture, so candidate network forms can be compared before a biological architecture is chosen. Neuromorphic and reservoir-computing studies already treat topology as a design variable \cite{Houben2026,Milisav2026}. Neurotopomorphic computing differs in its substrate, since the topology is physically patterned in living neurons. Circuits with different directional, modular and spatial organisations can be fabricated and then recorded through the same electrophysiological interface. The resulting architecture--communication--performance maps could identify topological motifs worth testing in silicon, hybrid or other neuromorphic substrates.

We propose that neuronal circuits designed for computation be measured at three levels: architecture, which sets the physical possibilities; the communication phenotype, which records how neuronal activity expresses them; and task performance, which shows which part of that activity a readout can use. Simulation allowed all three levels to be measured under matched stimulation and readout before fabrication, subject to the layout confounds noted in the Limitations. Patterned neuronal cultures are the next test, since biological variability, maturation and plasticity could each change how one level maps onto the next.

\subsection*{Limitations of the study}
Each circuit contained 15 Izhikevich neurons with fixed synaptic weights, and learning was confined to the readout. Short-term plasticity was evaluated only as a sensitivity analysis. Neuronal positions were geometric surrogates generated by force-directed and ring layouts, so edge lengths and delays follow from the layout rule, not from a fabricated device. The Microchannel Diode carries a circular embedding although its adjacency is a chain, so its path lengths and delays differ from those of the Sequential Chain. The source neuron was fixed at node 0 for Star, Sequential Chain and Microchannel Diode and selected by a reachability rule for the other six architectures. Architecture-level correlations rest on nine architectures. IC$^3$ weights were fixed before analysis and tested only under local perturbation. Post-stimulus response probability included spontaneous spikes.

\section*{STAR Methods}

\subsection*{Key resources table}

\begin{table}[h]
\centering
\footnotesize
\caption{Key computational resources used in this study.}
\begin{tabular}{p{0.24\textwidth}p{0.30\textwidth}p{0.20\textwidth}p{0.18\textwidth}}
\toprule
\textbf{Resource} & \textbf{Designation} & \textbf{Source} & \textbf{Identifier} \\
\midrule
Simulation and analysis code & Neurons-on-a-chip analysis package & This study & DOI: 10.5281/zenodo.22723783 \\
Numerical source data & Nine-architecture simulation dataset and figure-level data & This study & DOI: 10.5281/zenodo.22723783 \\
Python & Analysis environment & Python Software Foundation & v3.9.6 \\
NumPy & Numerical computation & NumPy developers & v1.26.1 \\
pandas & Data handling & pandas developers & v2.3.3 \\
SciPy & Scientific computation & SciPy developers & v1.11.3 \\
scikit-learn & Classification and regression readouts & scikit-learn developers & v1.6.1 \\
NetworkX & Graph generation and graph metrics & NetworkX developers & v3.2.1 \\
Matplotlib & Figure generation & Matplotlib developers & v3.8.1 \\
seaborn & Figure generation & seaborn developers & v0.13.2 \\
\bottomrule
\end{tabular}
\label{tab:key-resources}
\end{table}

\subsection*{Resource availability}
\paragraph{Lead contact.}
Further information and requests should be directed to the lead contact,
Michael Taynnan Barros (m.barros@essex.ac.uk).

\paragraph{Materials availability.}
This computational study did not generate new physical materials.

\paragraph{Data and code availability.}
The simulation code, analysis code, numerical source data, statistical
results, and figure-level data supporting this study are publicly available on
Zenodo at \url{https://doi.org/10.5281/zenodo.22723783}.

\subsection*{Methods}

\subsubsection*{Study design}
We compared nine candidate neuronal architectures under a fixed neuronal,
synaptic, and stimulation model. Principal analyses used $N=15$ neurons and 20
independently seeded simulated network instances per architecture. 
Stimulation frequencies, computational tasks, response windows, neurons,
transfer-entropy pairs, and stochastic trials were repeated within an instance.
Deterministic architectures kept the same connections across instances, while excitatory/inhibitory assignment, synaptic weights, edge velocities, and trial noise were resampled independently.

\subsubsection*{Network architecture}
\label{subsec:network-architecture-embedding}
 
Let
\begin{equation}
  \mathcal{A}_N
  =
  \left\{
    A\in \mathbb{R}_{\ge 0}^{N\times N}
    \;:\;
    A_{ii}=0 \ \forall i
  \right\}
\end{equation}
denote the space of non-negative directed connection templates on
$V=\{0,1,\ldots,N-1\}$.
An adjacency matrix $A=[A_{ij}]\in\mathcal{A}_N$ represents a directed edge
$i\to j$ whenever $A_{ij}>0$, with support
$E=\{(i,j):A_{ij}>0\}$.
Unless explicitly symmetrised, $A_{ij}$ need not equal $A_{ji}$.
Entries of $A$ are strictly non-negative: inhibition is never encoded by the
sign of a connection. Presynaptic cell identity instead routes each event to
the excitatory or inhibitory conductance, and the sign of the resulting current
follows from the corresponding reversal potential
(see Spiking neuron and synaptic dynamics).
 
For an architecture $\alpha$, the adjacency matrix was generated as follows.
All shortest-path quantities below are computed on the support graph $E$.
 
\begin{description}
  \item[All-to-All:]
  \begin{equation}
    A_{ij}=1 \quad \text{for all } i\neq j,
    \qquad
    A_{ii}=0.
  \end{equation}
 
  \item[Star:]
  \begin{equation}
    A_{0j}=A_{j0}=1 \quad \text{for } j=1,\ldots,N-1,
  \end{equation}
  with all other off-diagonal entries equal to zero.
 
  \item[Sequential Chain:]
  \begin{equation}
    A_{i,i+1}=1 \quad \text{for } i=0,\ldots,N-2,
  \end{equation}
  with all other entries equal to zero.
 
  \item[Microchannel Diode:]
  For $i=0,\ldots,N-2$,
  \begin{equation}
    A_{i,i+1}=1+B_i,
    \qquad
    B_i\sim \operatorname{Bernoulli}(0.8),
  \end{equation}
  and, independently,
  \begin{equation}
    A_{i+1,i}=0.3\,C_i,
    \qquad
    C_i\sim \operatorname{Bernoulli}(0.05).
  \end{equation}
  All other entries are zero, so the non-zero magnitudes take values in
  $\{0.3,\,1,\,2\}$.
 
  \item[Sparse and Directed random graphs:]
  Both architectures were generated as directed Erd\H{o}s--R\'enyi graphs in
which each ordered pair of distinct neurons was connected independently. The
two ensembles differed only in connection probability, with
$p_{\mathrm{Sparse}}=0.10$ and $p_{\mathrm{Directed}}=0.12$. Thus,

\begin{equation}
A_{ij}\sim\operatorname{Bernoulli}(p_{\alpha}),
\qquad i\neq j,
\end{equation}

where
\[
p_{\alpha}=
\begin{cases}
0.10, & \alpha=\mathrm{Sparse},\\
0.12, & \alpha=\mathrm{Directed\ Random}.
\end{cases}
\]

Diagonal entries were set to zero. We use the labels Sparse and Directed Random to distinguish the two directed-random networks, which differ in expected connection density.
 
  \item[Small-World:]
  Let $H_{\mathrm{SW}}=(V,E_{\mathrm{SW}})$ be an undirected Watts--Strogatz graph
  with neighbourhood degree $k=4$ and rewiring probability $p=0.10$.
  The directed adjacency is the symmetric embedding
  \begin{equation}
    A_{ij}=A_{ji}
    =
    \mathbf{1}\!\left\{\{i,j\}\in E_{\mathrm{SW}}\right\},
    \qquad i\neq j.
  \end{equation}
 
  \item[Clustered Ring:]
  This architecture requires $5\mid N$. Partition $V$ into five contiguous
  modules of equal size $\kappa=N/5$,
  \begin{equation}
    M_c=\left\{c\kappa,\;c\kappa+1,\;\ldots,\;(c+1)\kappa-1\right\},
    \qquad c=0,\ldots,4,
  \end{equation}
  and let $q(i)\in\{0,\ldots,4\}$ denote the index of the module containing $i$.
  Inter-module coupling is \emph{unidirectional} and cyclic, running only from
  module $c$ to module $c+1 \pmod 5$. For $i\neq j$,
  \begin{equation}
    p_{ij}
    =
    \begin{cases}
      0.75,
      & q(i)=q(j),\\[2mm]
      0.15,
      & q(j)\equiv q(i)+1 \pmod{5},\\[2mm]
      0,
      & \text{otherwise}.
    \end{cases}
  \end{equation}
  Edges were sampled independently over ordered pairs according to $p_{ij}$,
  while ensuring that each adjacent pair of modules remained connected by at least one forward edge in the cyclic direction,
  \begin{equation}
    \sum_{i\in M_c}\;\sum_{j\in M_{(c+1)\bmod 5}}
      \mathbf{1}\{A_{ij}>0\}
    \;\ge\; 1,
    \qquad c=0,\ldots,4.
  \end{equation}
  When independent sampling produced no forward edge for a given pair, a single
  uniformly chosen entry of that block was set to one.
 
  \item[Scale-Free:]
  Let $H_{\mathrm{SF}}=(V,E_{\mathrm{SF}})$ be an undirected Barab\'asi--Albert graph
  with attachment parameter $m=2$.
  The directed adjacency is the symmetric embedding
  \begin{equation}
    A_{ij}=A_{ji}
    =
    \mathbf{1}\!\left\{\{i,j\}\in E_{\mathrm{SF}}\right\},
    \qquad i\neq j.
  \end{equation}
\end{description}
 
\paragraph{Source selection.}
For Star, Sequential Chain, and Microchannel Diode, the source node was fixed at
$s=0$. For all other architectures, the source was selected by a lexicographic
rule on three graph quantities, computed on $G=(V,E)$.
Let $d_G(i,j)$ denote the shortest directed path length from $i$ to $j$, and let
\begin{equation}
  R^{+}(i)
  =
  \left\{
    j\in V\setminus\{i\}
    \;:\;
    d_G(i,j)<\infty
  \right\},
  \qquad
  r(i)=\left|R^{+}(i)\right|
\end{equation}
be the set and number of nodes reachable \emph{from} $i$. Downstream
accessibility is quantified by the out-closeness
\begin{equation}
  c^{+}(i)
  =
  \begin{cases}
    \displaystyle
    \frac{r(i)}
         {\sum_{j\in R^{+}(i)} d_G(i,j)}
    \cdot
    \frac{r(i)}{N-1},
    & r(i)>0,\\[5mm]
    0,
    & r(i)=0,
  \end{cases}
  \label{eq:out-closeness}
\end{equation}
where the second factor is the Wasserman--Faust correction for graphs in which
not all nodes are mutually reachable. Equation~\eqref{eq:out-closeness} uses
outgoing distances $d_G(i,\cdot)$ and measures how efficiently $i$
reaches the rest of the circuit. It is computed in the implementation as the
standard incoming-distance closeness applied to the reversed graph, which is
numerically identical.
The out-degree is
\begin{equation}
  k^{+}(i)
  =
  \sum_{j\in V}\mathbf{1}\{A_{ij}>0\}.
\end{equation}
The selected source is
\begin{equation}
  s
  =
  \begin{cases}
    0,
    & \alpha\in
      \{\text{Star},\text{Sequential Chain},\text{Microchannel Diode}\},\\[2mm]
    \displaystyle
    \operatorname*{arg\,max}_{i\in V}{}^{\mathrm{lex}}
    \left(
      r(i),\;
      c^{+}(i),\;
      k^{+}(i),\;
      -i
    \right),
    & \text{otherwise},
  \end{cases}
  \label{eq:source}
\end{equation}
where $\operatorname{arg\,max}^{\mathrm{lex}}$ compares the tuple entries in
order. The rule maximises reachable descendants first, then out-closeness, then
out-degree, and finally selects the lowest node index among remaining ties. It
prevents selection of a structurally isolated source when a reachable candidate
exists, without assuming that a topologically central node is functionally
optimal.
 
\paragraph{Physical embedding.}
Each graph was embedded in a square domain of side $L=650\,\mu\mathrm{m}$ with
margin $\rho=40\,\mu\mathrm{m}$. This sub-millimetre footprint was chosen to represent the spatial scale of engineered neuronal circuits used in microfluidic and MEA-based platforms, where neuronal compartments and axon-guiding structures commonly span tens to several hundred micrometres \cite{Habibey2022,Sato2023,Habibey2024}. 
\begin{equation}
  \Omega=[\rho,\,L-\rho]^2 .
\end{equation}
A layout is a map $X:V\to\Omega$, $i\mapsto x_i$. Raw layout coordinates
$y_0,\ldots,y_{N-1}\in\mathbb{R}^2$ were mapped into $\Omega$ by the per-axis
affine rescaling $\mathcal{N}$, applied independently to each Cartesian
component $d\in\{1,2\}$,
\begin{equation}
  \left[\mathcal{N}(y)\right]_{i,d}
  =
  \rho
  +
  \frac{y_{i,d}-\min_k y_{k,d}}
       {\max\!\left(\max_k y_{k,d}-\min_k y_{k,d},\,\varepsilon\right)}
  \left(L-2\rho\right),
  \qquad \varepsilon=10^{-12}.
  \label{eq:layout-normalisation}
\end{equation}
Because $\mathcal{N}$ rescales each axis separately, it is in general
anisotropic, e.g. circular raw layouts are mapped to axis-aligned ellipses inscribed
in $\Omega$.
 
The Sequential Chain used a linear midline layout, assigned directly without
rescaling,
\begin{equation}
  x_i
  =
  \left(
    \rho + i\,\frac{L-2\rho}{N-1},
    \;\frac{L}{2}
  \right),
  \qquad i=0,\ldots,N-1,\ N>1 .
\end{equation}
The Star used a centred radial layout, with raw coordinates
\begin{equation}
  y_0=(0,0),
  \qquad
  y_j=\left(\cos\theta_j,\,\sin\theta_j\right),
  \quad
  \theta_j=\frac{2\pi(j-1)}{N-1},
  \quad j=1,\ldots,N-1,
\end{equation}
followed by $X=\mathcal{N}(y)$.
Ring-like architectures used raw angular coordinates on the full node set,
\begin{equation}
  y_k=\left(\cos\tfrac{2\pi k}{N},\,\sin\tfrac{2\pi k}{N}\right),
  \qquad k=0,\ldots,N-1,
\end{equation}
followed by $X=\mathcal{N}(y)$. This ring-layout rule was applied to the Clustered Ring and, because of the architecture-specific layout mapping, also to the Microchannel Diode. The
Microchannel Diode carries a circular embedding even though its
adjacency is a chain, so its edge path lengths and propagation delays differ
systematically from those of the Sequential Chain despite the two architectures
having nearly identical connectivity. All remaining architectures were embedded using a seeded two-dimensional
Fruchterman--Reingold force-directed layout. Let

\begin{equation}
  Y=\operatorname{FR}(A;\xi),
\end{equation}

where $\operatorname{FR}(A;\xi)$ denotes the two-dimensional
Fruchterman--Reingold embedding of the adjacency matrix $A$ using a fixed seed
$\xi$. The resulting coordinates were then rescaled to the physical device
domain using Equation~\eqref{eq:layout-normalisation},

\begin{equation}
  X=\mathcal{N}(Y).
\end{equation}
 
The layout generator depended on the architecture and on the adjacency matrix:
\begin{equation}
  A\sim P_\alpha,
  \qquad
  X=\Lambda_\alpha(A;\xi).
  \label{eq:layout-map}
\end{equation}
Adjacency was defined by the prescribed circuit architecture. For some architectures, neuronal positions were generated from the network connectivity (Equation~\eqref{eq:layout-map}), so rewiring could change both the connections between neurons and their simulated spatial arrangement, which in turn altered propagation delays. We use the rewiring analysis to test how sensitive the results are to changes in the overall network configuration.
For an existing edge $(i,j)\in E$ the Euclidean separation is
\begin{equation}
  \ell_{ij}=\left\|x_i-x_j\right\|_2 ,
\end{equation}
which enters the total edge delay through the axonal path-length proxy, the
edge-specific conduction velocity and the fixed synaptic delay defined in
in Spiking neuron and synaptic dynamics. Soma placement was interpreted at the circuit scale as a geometric surrogate for edge-length and propagation-delay estimation.
 
\paragraph{Density, degree and the rewired control.}
For architecture
$\alpha$ the edge density is
\begin{equation}
  \varrho_\alpha=\frac{|E_\alpha|}{N(N-1)},
\end{equation}
and the in- and out-degrees are
\begin{equation}
  k_i^{\mathrm{in}}=\sum_{j\in V}\mathbf{1}\{A_{ji}>0\},
  \qquad
  k_i^{\mathrm{out}}=\sum_{j\in V}\mathbf{1}\{A_{ij}>0\}.
\end{equation}
These quantities were allowed to differ across primary architectures because the
complete patternable circuit was the engineering intervention. To isolate
organisation beyond the degree sequence, degree-preserving rewiring was analysed
separately. The rewired matrix $\widehat{A}$ was obtained by directed
double-edge swaps and satisfies both the degree constraints
\begin{equation}
  \sum_{j\in V}\mathbf{1}\{\widehat{A}_{ij}>0\}=k_i^{\mathrm{out}},
  \qquad
  \sum_{j\in V}\mathbf{1}\{\widehat{A}_{ji}>0\}=k_i^{\mathrm{in}},
  \qquad \forall\, i\in V,
\end{equation}
and the magnitude constraint that the multiset of non-zero entries is preserved,
\begin{equation}
  \left\{\!\!\left\{\widehat{A}_{ij}:\widehat{A}_{ij}>0\right\}\!\!\right\}
  =
  \left\{\!\!\left\{A_{ij}:A_{ij}>0\right\}\!\!\right\},
\end{equation}
with the reassignment of magnitudes to rewired edges drawn as a seeded uniform
permutation. Each swap requires four distinct nodes, so architectures whose
support admits no such swap are returned unchanged and are reported as having no
available rewired control.

\subsubsection*{Spiking neuron and synaptic dynamics} \label{subsec:spiking-neuron-synaptic}

Each node in the network was represented by an Izhikevich spiking neuron \cite{Izhikevich2003}. This model class has been used with stochastic drive and conductance-based synapses to reproduce activity from cortical neuronal cultures reconstructed from MEA recordings and, more recently, collective dynamics in engineered anisotropic neuronal cultures.\cite{Li2022,HoubenPLOS2025} We use it as a computationally efficient reduced model of a small neurons-on-a-chip circuit. Membrane dynamics followed the Izhikevich model with additive noise and a constant bias current,

\begin{align}
\mathrm{d}v_i &= \left(0.04v_i^2+5v_i+140-u_i+I_i^{\mathrm{syn}}+I_i^{\mathrm{bias}}+I_i^{\mathrm{stim}}\right)\mathrm{d}t+\alpha\,\mathrm{d}W_i(t),\\
\frac{\mathrm{d}u_i}{\mathrm{d}t} &= a_i(b_iv_i-u_i),
\end{align}

where $W_i$ were independent Wiener processes, $I_i^{\mathrm{bias}}=0.5$ and $\alpha=2.0$ in model units. Exactly 12 of 15 neurons were excitatory regular-spiking cells with
$(a,b,c,d)=(0.02,0.20,-65,8)$, and the remaining three were inhibitory
fast-spiking cells with $(0.10,0.20,-65,2)$. These RS and FS parameter
sets follow the canonical Izhikevich formulations \cite{Izhikevich2003}.
The 80:20 excitatory--inhibitory composition is consistent with
computational models of engineered and cultured neuronal networks
\cite{HoubenPLOS2025}. Excitatory and inhibitory reversal potentials
were set to 0 and $-80$\,mV, respectively, consistent with commonly used
AMPA- and GABA$_{\mathrm A}$-like conductance models. Axonal conduction \cite{Li2019}
velocities were constrained to 0.10--1.20\,m\,s$^{-1}$, spanning the
range measured in microchannel-coupled neuronal cultures, where more
than 95\% of recorded propagation velocities fell between 0.3 and
1.2\,m\,s$^{-1}$ \cite{Habibey2017}. The remaining bias, noise, and
synaptic-weight parameters were fixed model settings used consistently
across architectures. A spike at $v_i\geq30$\,mV caused $v_i\leftarrow c_i$ and $u_i\leftarrow u_i+d_i$.

Synaptic current was

\begin{equation}
I_i^{\mathrm{syn}}=g_i^{\mathrm E}(E_{\mathrm E}-v_i)+g_i^{\mathrm I}(E_{\mathrm I}-v_i),
\end{equation}

with $(E_{\mathrm E},E_{\mathrm I})=(0,-80)$\,mV and conductance time constants $(\tau_{\mathrm E},\tau_{\mathrm I})=(5,6)$\,ms. All outgoing events from an excitatory neuron incremented $g^{\mathrm E}$ and those from an inhibitory neuron incremented $g^{\mathrm I}$. For an edge $i\rightarrow j$, $w_{ij}=A_{ij}\omega_{\chi_i}Z_{ij}$, where $\omega_{\mathrm E}=\omega_{\mathrm I}=0.60$ and $Z_{ij}\sim\mathrm{LogNormal}(0,0.35^2)$. The mean weight per unit adjacency was therefore $0.60\,\mathbb{E}[Z_{ij}]=0.60\exp(0.35^2/2)\approx0.64$.

Conductances decayed exponentially between events,

\begin{equation}
\tau_{X}\frac{\mathrm{d}g_i^{X}}{\mathrm{d}t}=-g_i^{X},\qquad X\in\{\mathrm E,\mathrm I\},
\label{eq:cond-decay}
\end{equation}

integrated in closed form at each step as $g_i^{X}\leftarrow g_i^{X}\exp(-\Delta t/\tau_{X})$, after which all spikes arriving at that step were added, $g_i^{X}\leftarrow g_i^{X}+\sum_{j\in\mathcal A_i^{X}(n)}w_{ji}$, where $\mathcal A_i^{X}(n)$ is the set of $X$-type presynaptic events delivered to neuron $i$ at step $n$. Conductances used an exact exponential update. The decay-then-deliver order means an event arriving at step $n$ is not attenuated within step $n$.

The discrete noise current was $\alpha\xi_{i,n}/\sqrt{\Delta t}$, $\xi_{i,n}\sim\mathcal N(0,1)$, producing a voltage increment $\alpha\sqrt{\Delta t}\xi_{i,n}$; the recovery update used the post-voltage-update state. A hard 2.0\,ms absolute refractory clamp held $v_i=c_i$ for 16 time steps while recovery and conductance states continued to evolve. Excitatory parameters were $(U_0,\tau_{\mathrm{rec}},\tau_{\mathrm{fac}})=(0.50,\,800\,\mathrm{ms},\,0\,\mathrm{ms})$ and inhibitory parameters were $(0.05,\,200\,\mathrm{ms},\,600\,\mathrm{ms})$. Short-term plasticity was enabled only in the sensitivity analysis and was
implemented per edge using Tsodyks-type utilisation and resource variables
$(U,R)$, initialised at $U=U_0$ and $R=1$ \cite{Tsodyks1998}. Both variables were 
relaxed at each integration step, and synaptic release was evaluated at the
delayed spike-delivery time as $w_{ij}UR$, followed by resource depletion and,
for facilitating synapses, utilisation update. Excitatory synapses
($\tau_{\mathrm{fac}}=0$) expressed depression only.

Propagation delays were modelled from the physical geometry of each circuit by
combining axonal path length, heterogeneous conduction velocity, and a fixed
local synaptic delay. Soma-to-soma distance was multiplied by a tortuosity
factor of 1.15 to approximate the longer trajectory of neurite paths. The total delay for each edge was calculated as
axonal path length divided by conduction velocity plus a local synaptic delay
of 0.80\,ms and discretised to the simulation time step. Shortest physical path
delay was then defined as the minimum cumulative edge delay along all directed
source--target paths, providing a geometry-aware measure of communication
timing complementary to graph hop distance and measured neuronal response
latency \cite{Habibey2017,Gladkov2017}.

\subsubsection*{IC$^3$ theory}

IC$^3$ defines the Integrated Characterisation of Communication-Driven Computation. Its definition does not assume a particular neuronal model or measurement platform, and here we test one implementation. Let a neuronal system be $\mathfrak S=(G,\Theta)$, where $G$ is its physical or effective architecture and $\Theta$ contains its cellular and coupling parameters. Under input $u$ and stochastic realization $\xi$, the system produces a spatiotemporal neuronal state trajectory

\begin{equation}
\mathbf X=\Phi(G,\Theta,u,\xi).
\end{equation}

Three coupled domains are extracted from this trajectory and its architecture:

\begin{equation}
\mathbf q=\mathcal Q(\mathbf X,u),\qquad
\mathbf c=\mathcal C(\mathbf X,G,u),\qquad
\mathbf s=\mathcal S(G).
\end{equation}

$\mathcal Q$ describes \textit{neuronal dynamics}: the neuronal states generated, differentiated, and sustained under stimulation. $\mathcal C$ describes \textit{functional communication} through information exchange, propagation, and recruitment. $\mathcal S$ describes the \textit{structural support} that constrains both domains. After domain-specific normalisation, these quantities form the IC$^3$ configuration profile

\begin{equation}
\mathbf z_{\mathrm{IC3}}=
\left[
\mathcal N_Q(\mathbf q),
\mathcal N_C(\mathbf c),
\mathcal N_S(\mathbf s)
\right].
\label{eq:ic3-profile}
\end{equation}

A measurement model $\mathcal M$ specifies the observables, reference scales, orientation, and aggregation used in a given application. The general scalar index is

\begin{equation}
\mathrm{IC3}_{\mathcal M}=\mathcal A_{\mathcal M}(\mathbf z_{\mathrm{IC3}}).
\label{eq:ic3-general}
\end{equation}

We use the decomposable
weighted form

\begin{equation}
\mathrm{IC3}_{\mathcal M}
=\sum_{d\in\{Q,C,S\}}\alpha_d
\sum_{m\in\mathcal M_d}\omega_{dm}\,
\mathcal N_{dm}(x_{dm}),
\quad
\sum_d\alpha_d=1,
\quad
\sum_{m\in\mathcal M_d}\omega_{dm}=1.
\label{eq:ic3-weighted-general}
\end{equation}

Here, $\alpha_d$ determines the contribution of each domain, $\omega_{dm}$
determines the contribution of each observable within that domain, and
$\mathcal N_{dm}$ maps each observable onto a common dimensionless scale.
This formulation preserves the three-domain structure of IC$^3$ while allowing
the specific observables and weighting scheme to be adapted to the measurement
platform and scientific question. Equations~\eqref{eq:ic3-general} and \eqref{eq:ic3-weighted-general} quantify aggregate expression of the measured domains. The neurons-on-a-chip implementation used in
this study is defined in the next subsection.

The IC$^3$ profile is defined independently of task performance, although it remains conditional on architecture, input, and operating state. For task $\tau$, input encoding $u_\tau$, and readout $r_\tau$, performance is a task-specific projection

\begin{equation}
P_\tau=\Pi_\tau(\mathbf z_{\mathrm{IC3}},u_\tau,r_\tau).
\label{eq:ic3-task-projection}
\end{equation}

Two tasks may depend on different properties of the same configuration profile, and the same architecture may be useful for one transformation but not another.

\subsubsection*{IC$^3$ implementation for neurons-on-a-chip}

For each of the nine neuronal architectures, IC$^3$ combined eight observables
across three domains: neuronal dynamics, functional communication, and
structural support. The neuronal dynamics domain $\mathcal Q$ comprised
mean spike entropy, effective dimensionality, mean inter-spike-interval (ISI)
coefficient of variation, burst-like fraction, and regular-spiking fraction.
The functional-communication domain $\mathcal C$ comprised mean pairwise
mutual information and  transfer entropy (TE), while the
structural-support domain $\mathcal S$ was represented by a graph metric
combining directed degree and core number.

Each observable $x_m$ was mapped to a dimensionless component

\begin{equation}
z_m=\operatorname{clip}\left(\frac{x_m}{s_m},0,1\right),
\end{equation}

and the scalar IC$^3$ index was calculated as

\begin{equation}
\mathrm{IC3}_{\mathcal M}
=
\sum_{m=1}^{8}\omega_m z_m .
\label{eq:ic3-noc}
\end{equation}

IC$^3$ was weighted to reflect the progression from neuronal activity to
network communication and, finally, to the structural substrate supporting
both. The weights were a design choice fixed before analysis. We assigned the largest contribution to neuronal dynamics because this domain captures the activity states through which computation is expressed.
Within this domain, spike entropy and effective dimensionality describe
complementary population-level properties of state diversity and dimensional
structure, and were given greater emphasis than the three
single-neuron firing-regime descriptors: ISI variability, burst-like fraction,
and regular-spiking fraction. Functional communication formed the second
domain, with equal contributions from mutual information and 
transfer entropy to capture statistical dependence and directed information
transfer, respectively. Structural support provided the remaining contribution
through the graph-based architectural metric, representing the physical
constraint within which the observed neuronal dynamics and communication
emerge.

This hierarchy gives domain weights of 0.60 for neuronal dynamics, 0.30 for
functional communication, and 0.10 for structural support, corresponding to
component weights of
$(0.15,0.15,0.10,0.10,0.10,0.15,0.15,0.10)$ for spike entropy,
effective dimensionality, ISI variability, burst-like fraction,
regular-spiking fraction, mutual information,  TE, and structural
support, respectively. The corresponding reference scales were $(1,15,2,1,1,1,0.1,100)$. 

Mutual information, effective dimensionality, and TE were calculated from
10\,ms binned spike trains, while spike entropy was calculated at the
0.125\,ms simulation time step. First-order TE,
$I(X_{t-1};Y_t\mid Y_{t-1})$, was estimated across all directed neuron pairs
using 200 circular-shift surrogates to correct finite-sample bias
\cite{Schreiber2000}.  TE in bits contributed to IC$^3$, while
target-entropy-normalised TE was reported separately as a directed
communication measure.

Sensitivity analyses evaluated equal weighting, leave-one-component-out
variants, and 500 local perturbations of the IC$^3$ weights. Task-matched
IC$^3$ values were calculated from the same neuronal trajectories used for
frequency decoding, temporal-order discrimination, and fading-memory
analysis, allowing network configuration and computational performance to be
compared under matched operating conditions.

\subsubsection*{Structure-to-function analysis}

For each source--target pair, structural reachability, hop distance, and shortest
physical path delay were computed on the directed graph, excluding the stimulated
source from downstream statistics. Post-stimulus response probability was defined
as the fraction of stimulus onsets followed by at least one target spike within
50\,ms. Sensitivity analysis used windows of 10--150\,ms. This measure includes
spontaneous activity. Functional recruitment was defined at $R\geq0.5$, with
0.25 and 0.75 evaluated as sensitivity thresholds. The conversion from structural
reachability to functional recruitment was quantified as

\begin{equation}
\etaSF =
\frac{N_{\mathrm{functionally\ recruited}}}
     {N_{\mathrm{structurally\ reachable}}}.
\end{equation}

Response latency was measured from stimulus onset to the first target spike within
the response window. Neuron-level latency was summarised by the median, and
latency jitter by the sample standard deviation across valid responses. The
network communication phenotype combined post-stimulus response probability,
latency, upper-tail firing rate, latency jitter, ISI variability, and spike
expenditure.

\subsubsection*{Computational tasks and readouts}
\label{subsec:tasks-readouts}

We evaluated three complementary computational properties of each neuronal
architecture: discrimination of nearby input frequencies, discrimination of
temporal order, and retention of recent input history.
\textit{Frequency decoding} tests whether small differences in periodic input generate
separable population states. \textit{Temporal-order} discrimination tests whether the
network preserves the sequence of spatially distinct inputs. \textit{Fading memory} measures how much information about earlier inputs remains in the current neuronal state.
Network dynamics and model parameters remained fixed throughout these assays and
learning was confined to readouts.

Close-frequency decoding and fading-memory stimulation were applied to the architecture-specific source neuron defined in Network architecture.

Close-frequency decoding used four stimulation classes at 8.5, 10.0, 11.5,
and 13.0\,Hz to create a narrow-band temporal discrimination problem around the
operating regime used in the main simulations. Each class contained five trials
per network instance. Stimulation strength was fixed at $I_{\mathrm{stim}}=6$ across the principal
simulations, providing a common input operating point for comparisons across
architectures and stimulation conditions. Each trajectory lasted 1200\,ms. Population activity
was represented by dividing each trajectory into eight equal temporal windows
and concatenating spike counts from all 15 neurons.

Temporal-order discrimination tested whether the same network could distinguish
the sequence of two spatially separated inputs. Equal pulse amplitude and duration kept the stimulation energy matched
between classes, while the 75--125\,ms inter-pulse interval allowed the network
state to evolve between two inputs. The first pulse occurred between
150 and 250\,ms, and five trials per class were generated for each network
instance over 2000\,ms trajectories. Population activity was encoded using the
same eight-window spike-count representation used for frequency decoding.

For both classification tasks, neuronal features were aligned across
independently seeded networks according to functional role. The stimulated source was assigned the first position. In the temporal-order task, stimulation sites A and B were kept in their designated positions, while the remaining neurons were ordered by directed distance and node identity. Classification used
$L_2$-regularised multinomial logistic regression. Feature standardisation and
selection of $C\in\{0.1,1,10\}$ were performed within the training data of each
outer leave-one-network-instance-out fold using five-fold grouped
cross-validation. 

For fading memory, each network received a binary telegraph input on a
10\,ms grid with switching probability 0.15 and $I_{\mathrm{stim}}=6$. Three 2000\,ms trajectories were
generated per network instance. The neuronal state was represented by
per-neuron spike counts over the preceding 20\,ms and sampled every 10\,ms after
a 500\,ms washout period. Ridge-regression readouts reconstructed input states
25, 50, 100, and 200\,ms in the past. The regularisation parameter
$\lambda\in\{0.01,0.1,1,10,100\}$ was selected by grouped inner
cross-validation. Held-out $R^2$ equals zero when the readout predicts only the mean input. As a reference for the autocorrelation of the telegraph input, we also computed the $R^2$ obtained by predicting each lagged input from the current input state alone.

Classification accuracy $A$ was normalised relative to empirical chance as

\begin{equation}
S=\frac{A-C_{\mathrm{chance}}}{1-C_{\mathrm{chance}}},
\end{equation}

where $C_{\mathrm{chance}}$ was estimated by training-label permutation through
the same nested readout pipeline. Fading-memory performance used held-out $R^2$ directly.
Accordingly, $S=0$ represents baseline-level prediction, $S=1$ perfect
prediction, and $S<0$ performance below baseline.

We then examined how task-relevant information was distributed across each
circuit. Three nested feature sets were compared: $H_0$, containing the directly
stimulated neuron or neurons; $H_0+H_1$, which additionally included first-hop
neurons; and FULL, containing all aligned neuronal features. The incremental
decoding contributions were

\begin{equation}
\Delta P_{H_1}=S_{H_0+H_1}-S_{H_0},
\qquad
\Delta P_{\mathrm{FULL}}=S_{\mathrm{FULL}}-S_{H_0}.
\end{equation}

These contrasts quantify how expanding the readout from the stimulated nodes to
progressively larger regions of the network changed held-out decoding
performance.

A complementary rewiring analysis examined the sensitivity of task performance
to network organisation while preserving each node's in-degree and out-degree.
Degree-preserving directed double-edge swaps generated the rewired networks, and
the corresponding contrast was

\begin{equation}
\Delta P_{\mathrm{rewire}}
=
S_{\mathrm{native}}-S_{\mathrm{rewired}}.
\end{equation}

The network rewiring left the All-to-All and Star networks unchanged, so these architectures were excluded from the comparison.
For the remaining architectures, $\Delta P_{\mathrm{rewire}}$ measures how much performance changes after rewiring.

\subsubsection*{Simulation design and robustness}

\paragraph{Stimulation and replication.}
The main dynamics sweep used a 50\%-duty unipolar square-wave input with
$I_{\mathrm{stim}}=6$ at 0.5, 1, 2, 5, 10, 20 and 40\,Hz. Each trajectory
lasted 2000\,ms and three stochastic trials were generated for every condition.
The 0.5- and 1-Hz conditions contained approximately one and two
stimulation cycles, respectively, and are interpreted as finite-window
operating points. Structure--function analysis used the 5\,Hz condition, while
communication-phenotype analysis used 2, 5, 10 and 20\,Hz. These analyses
reused the corresponding dynamics-sweep trajectories.

\paragraph{Input and readout assignment.}
The architecture-specific source $s$ defined above was used as the stimulation
site for the single-input assays. For Star, Sequential Chain, and Microchannel
Diode networks, this corresponded to node 0; for the remaining architectures,
the source was selected independently for each network instance using
Equation~\eqref{eq:source}. In the temporal-order task, stimulation site A was the
architecture-specific source and site B was the physically most distant neuron
from A in the embedded circuit. No dedicated output neuron was imposed. Structure--function analyses treated
all downstream neurons except the stimulated source as potential targets.
For the computational assays, the full-network readout used the activity of all
15 neurons. Restricted H0 and H0+H1 readouts were used only to quantify the
contribution of the directly stimulated and first-hop populations.

\paragraph{Robustness analyses.}
Finite-size robustness was evaluated at $N\in\{10,15,25\}$ using 10
independently seeded instances per architecture and size. Device dimensions
were scaled by $\sqrt{N/15}$ to maintain approximately constant neuronal
density. Finite-size and short-term-plasticity comparisons used 10\,Hz
stimulation. Short-term plasticity was evaluated by comparing fixed OFF and ON
parameterisations without retuning.
IC$^3$ robustness was further assessed using alternative component weighting
and leave-one-component-out analyses, while TE sensitivity was evaluated across
alternative bin widths. Architecture rankings for IC$^3$ and task performance
were analysed separately. For the finite-size analysis, the effective-dimensionality scale remained
$s=15$ for all three network sizes. 

\subsubsection*{Computational environment and reproducibility}
Analyses were executed under Python 3.9.6 with NumPy 1.26.1, pandas 2.3.3,
SciPy 1.11.3, scikit-learn 1.6.1, NetworkX 3.2.1, Matplotlib 3.8.1, and seaborn
0.13.2. Randomness used NumPy PCG64 generators with deterministic
SHA-256-derived integer seeds. The archived package contains the
code and data required to reproduce the reported analyses and figures
\cite{BarrosZenodo2026}.

\subsection*{Quantification and statistical analysis}
Independent network instances were treated as the unit of replication, with
stochastic trials averaged within each instance and condition. Effects of
architecture, stimulation frequency, and their interaction were assessed using
Gaussian linear models with CR1 cluster-robust covariance by network instance.
Overall effects were evaluated using cluster-Wald tests, with
Benjamini--Hochberg correction applied where multiple related endpoints were
tested.

Task-specific architecture effects were evaluated across the nine architectures
and summarized using $\eta^2$. Comparisons between nested readouts and between
native and rewired networks used paired analyses across matched network
instances with confidence intervals where applicable.

Associations between IC$^3$,  TE, and task performance were analysed
separately at architecture and within-architecture levels. Architecture-level
analyses used architecture means with confidence intervals and exact
permutation tests. All tests were two-sided unless stated
otherwise. All analysis outputs used for the reported statistics are included in the archived study package \cite{BarrosZenodo2026}.



\section*{Author contributions}

M.T.B.: Conceptualisation, Methodology, Software, Validation, Formal analysis,
Investigation, Resources, Data curation, Writing the original draft, Review \& editing, and Visualisation.

\section*{Declaration of interests}

The author declares no competing interests.

\section*{Ethics statement}

This computational study used no human participants, human tissue, animals, or identifiable personal data. Ethical approval and informed consent were not applicable.

{\raggedright

}







\end{document}